# Flapping-pattern change in small and very small insects

Yu Zhu Lyu, Hao Jie Zhu and Mao Sun*

Institute of Fluid Mechanics, Beihang University, Beijing, China

([* m.sun@buaa.edu.cn](* m.sun@buaa.edu.cn))

Medium and large insects in normal hovering have horizontal, planar up- and downstrokes[1-4]. The lift of the two half-strokes, generated by the leading-edge vortex[5-10], provides the weight-supporting vertical force. But for small insects (wing length $R$ less than about 4 mm and Reynolds number $Re$ very low, about 80 to 10)[11], because of the large effect of air viscosity (as $Re$ becomes very low, moving in air is like in oil), sufficient vertical force could not be produced if using the above wing kinematics[12,13]. Small insects must use different flapping mode. Here, through analyzing flight data from our recent studies on a relatively-large small insect (fruitfly: $R\approx3$ mm, $Re\approx80$)[14] and a very small insect (wasp: $R\approx0.5$ mm, $Re\approx10$)[15], we put forward a hypothesis on how the flapping pattern will change: as insect-size or $Re$ decreasing, a deeper and deeper U-shape upstroke will be used to overcome the viscous effect. And we test this hypothesis by measuring the wing kinematics for species of different sizes to obtain data for $Re$ ranging from 80 to 10 and by computing the aerodynamic forces. The data and computation support our hypothesis: the planar upstroke changes to U-shape upstroke which becomes deeper as size or $Re$ becomes smaller; for relatively-large small insects, the U-shape upstroke produces a larger vertical force than a planar upstroke by having a larger wing velocity, and for very small insects, the deep U-shape upstroke produces a large transient drag that points almost upwards by fast downward acceleration of the wing, providing the required vertical force.

About half of the existing winged insect species are of small size (wing-length $R\approx0.5$-4 mm)[11]. But much of what we know about the biomechanical mechanisms of insect flight is derived from studies on medium and large insects ($R\approx5$-50 mm) insects[5-10], such as flies, honey-bees and hawkmoths. The non-dimensional parameter representing the effect of air viscosity is the Reynolds number (*Re*): lower *Re* means the wing moving in a more viscous flow. For the flapping wing of an insect, *Re* is approximately proportional to the square of *R* (*Re* is defined using the mean chord-length $c_m$ and the mean wing speed $U=2\Phi f r_2$ where $\Phi$ is the stroke aptitude, *f* the stroke frequency and $r_2$ the radius of gyration of wing). Thus the wings of the small insects operate at very low *Re*, on the order of 80-10. At this range of *Re*, moving in the air is like in oil.

Medium and large insects in normal hovering beat their wings approximately in a horizontal plane (Fig. 1a)[1-4] and the wings operate at Reynolds number (*Re*) about 100-3500. During the downstroke or upstroke, a lift, and a drag that is a little smaller, are produced (Fig. 1a). The lift provides the weight supporting vertical force; the drag in the downstroke cancels out that in the upstroke and the flapping-cycle mean horizontal force is zero. The aerodynamic forces are generated mainly by the leading-edge vortex (LEV) that attaches to the wing in the entire up- or downstroke, which is referred to as the delayed-stall mechanism[5-10]. However, if the small insects flap their wings as their larger counterparts, sufficient aerodynamic force cannot be produced because of the very strong viscous effects: the LEV is significantly defused and little lift can be generated, while the drag is very large[12,13]. The small insects must have used different wing kinematics and aerodynamic mechanisms from those of the medium and large insects.

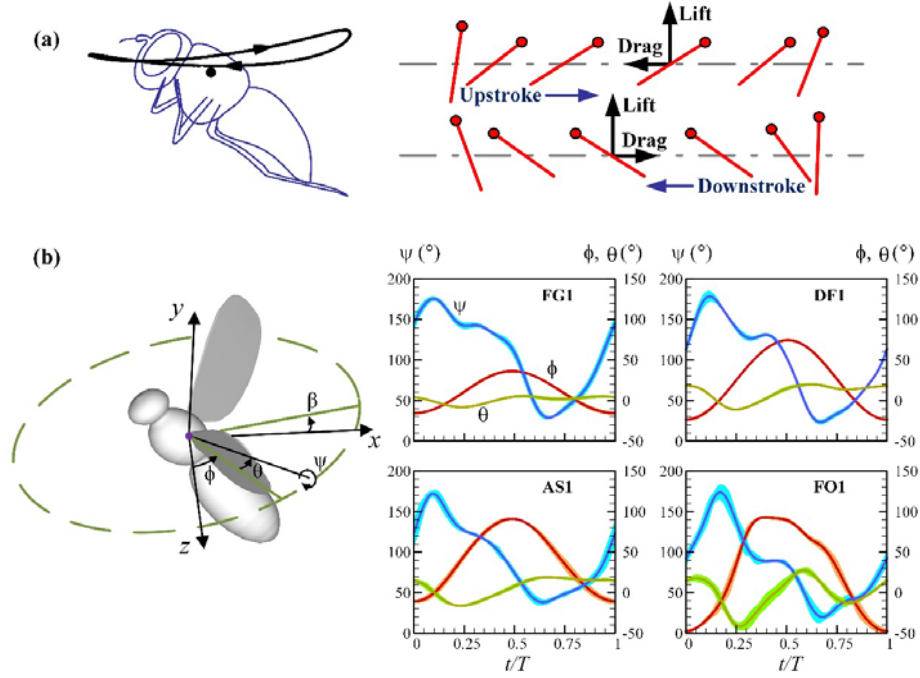

**Figure 1**. **(a)** Left: the wing-tip trajectory (projected onto the symmetrical plane) of a dronefly *Eristalis tenax* (ET)[4], a relatively large insect; right: the motion of a section of the wing. **(b)** Left: reference frame and Euler angles defining the wing kinematics: ($x$, $y$, $z$) are coordinates in a system with its origin at the wing root and with $x$-axis points horizontally backwards and $z$-axis points vertically upwards and $y$-axis points to the leftt of the insect, $\phi$ is the positional angle (in the stroke plane), $\psi$ the pitch angle, $\theta$ deviation angle, and $\beta$ the stroke-plane angle; right: measured Euler angles of biting-midge FG1 (mean±s.d.; $n$=30 wingbeats), biting-midge DF1 (mean±s.d.; $n$=30 wingbeats), gall midge AS1 (mean±s.d.; $n$=6 wingbeats) and thrip FO1 (mean±s.d.; $n$=6 wingbeats). $T$, stroke period.

From our recent studies on hovering of a relatively-large small insects, fruitfly *Drosophila virilis* (DV) ($R$≈3 mm, $Re$≈80)[14] and a very small insect, wasp *Encarsia Formosa* (EF) ($R$≈0.5 mm, $Re$≈10)[15], we observe that the wasp has a very deep U-shape upstroke and the fruitfly also has a U-shape upstroke but much shallower, and that the deep U-shape upstroke can generate a large transient-drag that pointed almost upwards, enhancing the vertical force production. Based on this observation, we put forth a hypothesis that as insect-size or $Re$ decreasing, deeper and deeper U-shape upstroke would be used to overcome the viscous effect. We test this hypothesis by measuring the wing kinematics of more species of different sizes and obtained data for *Re* ranging from about 80 to 10. A few years earlier, we measured the wing motion of vegetable leafminers *Liriomyza sativae* (LS) ($R$≈1.5 mm and $Re$≈40)[16]. Here the wing kinematics of four more species are measured: biting-midges *Forcipomia gloriose* (FG) and *Dasyhelea flaviventris* (DF), gall-midge *Anbremia* sp. (AS) and Thrip *Frankliniella occidentalis* (FO). Their *Re* is 30, 24, 17 and 14, respectively, and is between those of LS (40) and EF (10). Fig. 1b shows the measured Euler angles of wing in four individuals, each from one of the four species (Supplementary Videos 1-4; the videos can also be seen at: msun.buaa.edu.cn). For each of the four species, data of another four individuals (Supplementary Table 1) were also measured; within a species, the results are similar (Supplementary Fig. 1).

Using data in Fig. 1b, stroke diagrams showing the flapping mode of the four insects are plotted in Fig. 2 (those of small insects DV[14], LS[16] and EF[15] and a large insect, dronefly *Eristalis tenax* (ET)[4], are also included). The results support our hypothesis: as size or *Re* decreasing, the insect has a deeper and deeper U-shape upstroke (Fig. 2b-h). For the two very small insects, thrip FO and wasp EF, the downstroke is also U-shaped, but a much shallower one (Fig. 2g, h), which is the result of the 'fling' motion discovered by Weis-Fogh[1] (described in detail elsewhere[15,17-20]). Within each of the species, all the individuals have the same flapping pattern (Supplementary Fig. 2).

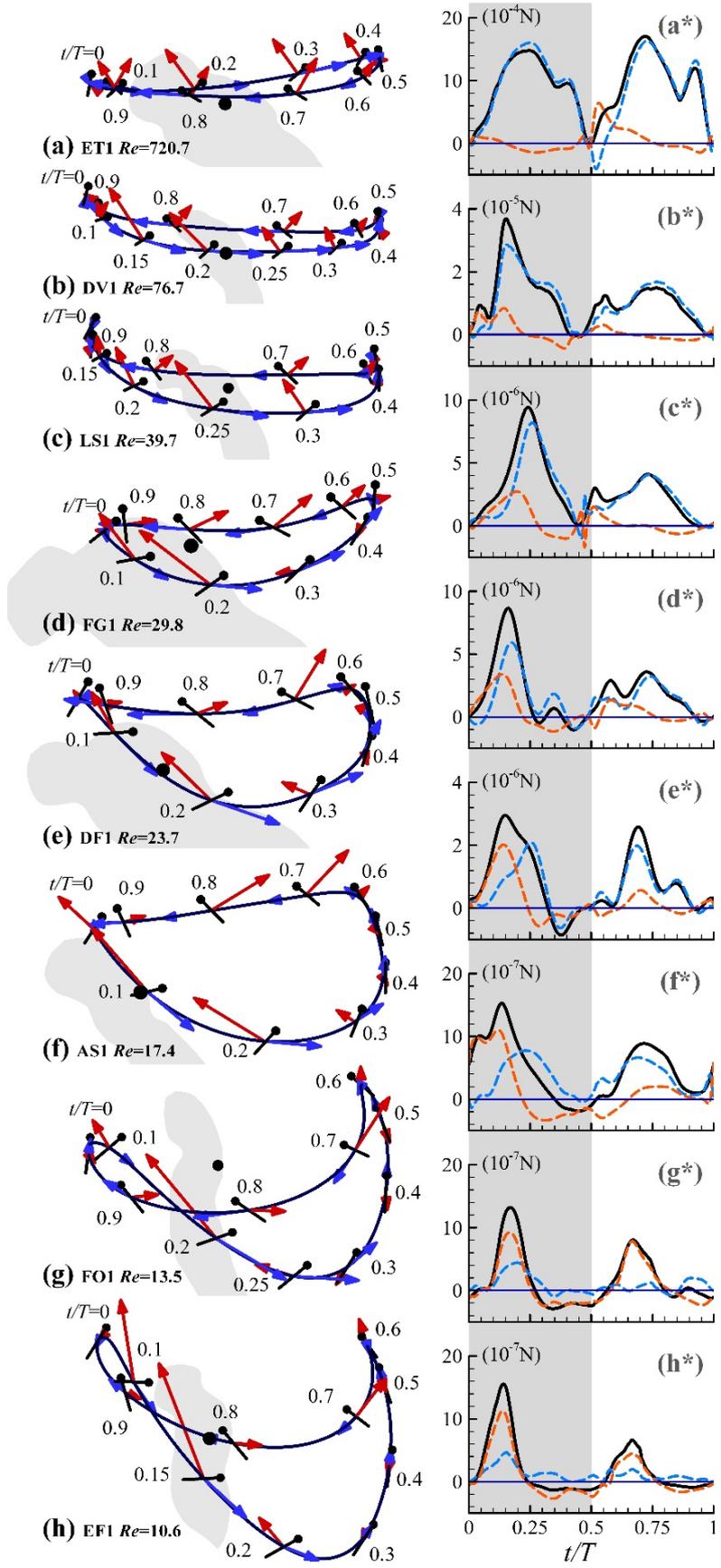

**Figure 2. (a) through (h)**: stroke diagrams show the wing motions of the insects; the solid curve indicates the wing-tip trajectory (projected onto the *x-y* plane); black lines indicate the orientation of the wing at various times in one stroke cycle, with dots marking the leading edge; a black dot defines the wing-root location on the insect body; the blue arrow represents the velocity of the wing at the radius of gyration; the red arrow represents the total aerodynamic force of the wing. **(a\*) through (h\*)**: vertical forces in one stroke cycle (black line); contribution by the lift (blue line) and that by drag (orange line); gray background indicates the upstroke.

In the early part of the U-shape upstroke ($t/T$≈0-0.2), the wing accelerates downwards and backwards with wing surface almost horizontal and at a rather large angle of attack; the smaller the insect or the lower the *Re*, the larger the acceleration (see the change in wing speed in Fig. 2b-h; here, the velocity at the radius of gyration of wing is used to represent the velocity of the wing). In the later part of the U-shape upstroke, generally the wing moves slower (Fig. 2e-h); as *Re* becomes very small, the wing moves almost vertically upwards with the wing surface vertical and the angle of attack close to zero (Fig. 2e-h).

To assess how the weight-supporting force is generated, the flow and forces on the wings were computed using an experiment-data validated flow solver. Fig. 2a\*-h\* give the computed vertical forces ($F_V$) for the eight insects shown in Fig. 2a-h (the corresponding horizontal forces are given in Supplementary Fig. 3; the forces for the other individuals of each species given in Supplementary Fig. 4). For the large insect ET (Fig. 2a\*), $F_V$ in the upstroke ($t/T$=0-0.5) is similar to that in the downstroke ($t/T$=0.5-1). But for the small insects (Fig. 2b\*-h\*), the U-shape upstroke produces a large $F_V$ peak, while the planar downstroke produces a relatively small $F_V$. In general, as size or *Re* deceasing, the $F_V$ peak in the U-shape upstroke becomes higher and narrower. For the two very small insects (FO and EF), a smaller $F_V$ peak is also produced at the early downstroke (Fig. 2g* and h*, $t/T$=0.55-0.7) by the 'fling' motion aforementioned. The vertical and horizontal forces of a wing come from its lift and drag, which are the components of the total aerodynamic force that are perpendicular and parallel to the wing-velocity, respectively. For the large insect ET (Fig. 2a*), almost all the vertical force is contributed by the lift. For the relatively large small insects DV and LS (Fig. 2b* and c*), vertical force is mainly (about 90%) is contributed by the lift. As insects become smaller (Fig. 2d*-h*), the drag has more and more contribution; for the two smallest insects, FO and EF, about 70% vertical-force is from the drag.

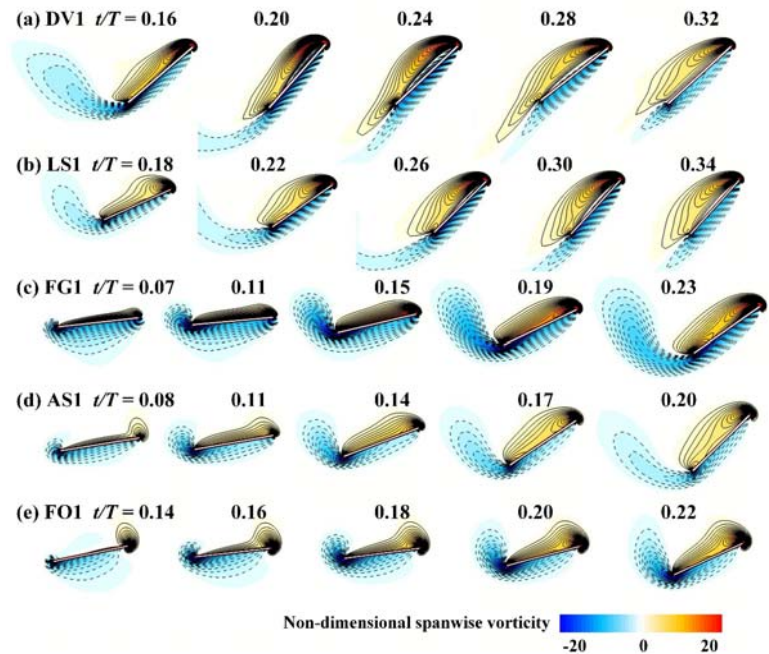

**Figuere. 3**. Non-dimensional spanwise vorticity contours in the section at the radiius of gyration, at verious times during the U-shape upstroke. Vorticity is nondimensionalized by $U/c_m$; solid and dashed lines denote the anticlockwise and clockwise vorticity, respectively. The magnitude of the non-dimensional vorticity at the outercontour is 2 and the contour interval is 1.

To explain the large force in the U-shape upstroke, vorticity-fields are plotted for the period in which the large aerodynamic force is produced. We first consider DV and LS, whose size is relatively large and $Re$ relatively high (about 80-40). Fig. 3b shows the vorticity plots of LS1 in the period of $t/T$=0.18-0.34: a LEV attaches and moves with the wing, indicating that the force is produced by the delayed-stall mechanism. It can be shown that the planar downstroke also use the delayed-stall mechanism to produce the force. The reason for the U-shaped upstroke producing a larger force than the planar downstroke is that it has a larger wing velocity (see Fig. 2c). The forces of DV can be similarly explained. Next we consider the smaller insects (Figs. 3c-e). As an example, we look at the vorticity plots of FO (Fig. 3e): during the very short period ($t/T$=0.14-0.22), counter clockwise vorticity is continuously produced around the leading edge of the wing and clockwise vorticity around the trailing edge. This would result in a large time rate of change in the first moment of vorticity, giving the large aerodynamic force[21]. This force producing mechanism is called as 'rowing mechanism'[15,22]: the wing accelerates fast from zero velocity at a very high angle of attack, producing a large transient drag (here the drag points almost upwards, giving the large vertical force). The same is true for FG, DF, AS and EF.

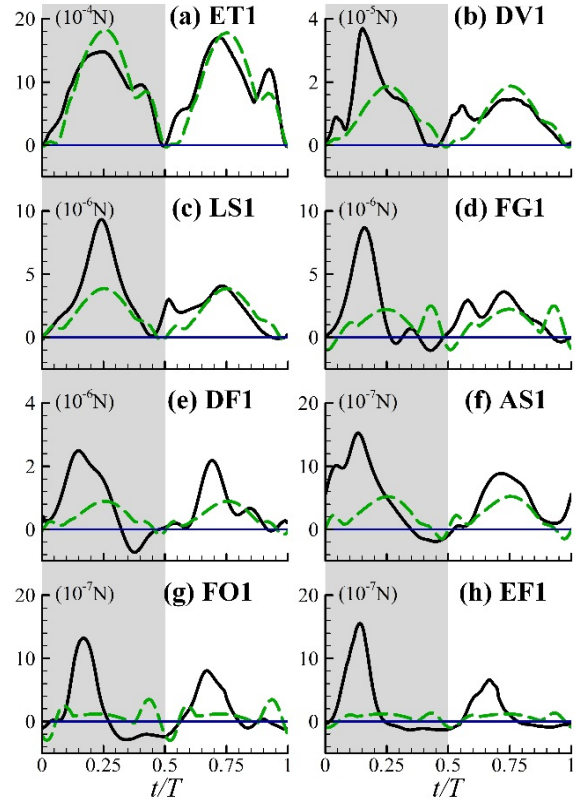

**Figuere. 4. (a) through (h)**: vertical force produced if the upstroke was planar (green dashed line), compared with that using the real wing kinematics which has U-shape upstroke (black line).

To show the advantage of having a U-shape upstroke for the small insects, we made test calculations in which both the down- and upstrokes were planar and horizontal, like those of the larger insects, while $Re$ being kept the same. The computed vertical forces for the eight insects are shown in Fig. 4a-h (the corresponding horizontal forces are given in Supplementary Fig. 5). It is seen that the planar upstroke produce much less vertical force than the U-shape upstroke and it is more so as $Re$ becomes smaller. For the fruitfly DV ($Re\approx77$), the mean vertical force produced by the U-shape upstroke is 1.43 times of that by the planar upstroke. For LS ($Re\approx40$), FG ($Re\approx30$), DF ($Re\approx24$), AS ($Re\approx17$), FO ($Re\approx14$) and EF ($Re\approx10$), the corresponding numbers are 1.82, 1.75, 1.92, 2.27, 2.27, and 4.17, respectively. This shows that the small insects use the novel motion, the U-shape upstroke, to overcome the problem of insufficient lift at very small Reynolds number, encountered by the commonly used flapping kinematics.

Taken together, our findings show that the small insects change their flapping mode to solve the low-$Re$ problem: The planar upstroke changes to U-shape upstroke, and as size or $Re$ becomes smaller, a deeper U-shape upstroke is employed; for relatively-large small insects ($Re$ about 80-40), the U-shape upstroke produces a larger vertical force than a planar upstroke by having a larger wing velocity, and for very small insects ($Re$ below about 40), the deep U-shape upstroke produces a large transient drag that points almost upwards by fast downward acceleration of the wing, providing the required vertical force.

## Methods

**Insects.** Thrips (FO) were acquired from the Laboratory of Institute of Vegetables and Flowers, Chinese Academy of Agricultural Sciences, which were descendents of wild-caught *Frankliniella occidentalis*. Biting-midges *Forcipomia gloriose* (FG) and *Dasyhelea flaviventris* (DF), gall midges *Anbremia* sp. (AS) were netted in a suburb of Beijing in June to August 2017.

**High-speed filming.** The near-hover flights of the small insects in transparent flight chambers were filmed using three orthogonally aligned synchronized high-speed cameras (FASTCAM Mini UX100, Photron Inc., San Diego, CA, USA) mounted on an optical table (Supplementary Fig. 6a). The size of the flight chamber is 50×50×50

mm$^3$ for biting-midge *Forcipomia gloriose* (FG) and 34×34×34 mm$^3$ for biting-midge *Dasyhelea flaviventris* (DF), gall-midge *Anbremia* sp. (AS) and Thrip *Frankliniella occidentalis* (FO). Each camera was equipped with a 60 mm micro-Nikkor lens and 12 mm extension tube. For FG, the cameras were set to 10,000 (or 10500) frames per second (resolution 1280×496 or 1280×472) pixels and shutter speed to 20 μs; for DF, AS and FO, to 8,000 frames per second (resolution 1280×624 pixels) and shutter speed to 20 μs. Each camera view was backlit using a 50 W integrated red light emitting diode (LED; luminous flux, 4000 lm; wavelength, 632 nm) and two lenses were used to make the light uniform. The synchronized cameras were manually triggered when the insect was observed to fly steadily in the filming area (approximately 6×6×6 mm$^3$) which represented the intersecting field of views of the three cameras. The experiment was performed at temperature 25-27˚C and relative humidity 50-60%.

**Kinematics reconstruction.** The orthogonally aligned cameras were calibrated by using a flat glass panel with a high accuracy black-and-white checkerboard pattern printed on it. The calibration gave the intrinsic and extrinsic parameters of each camera which determined the transform matrix of the camera[4,23].

The method used to extract the 3D body and wing kinematics from the filmed data was developed in previous works of our group[4,23,15]. The body and wings were represented by models (Supplementary Fig. 6b): the model of the body was two lines perpendicular to each other, which were the line connecting the head and the end of the abdomen and the line connecting the two wing hinges (Supplementary Fig. 6b);

the model of a wing was the outline of the wing obtained by scanning the cut-off wing (Supplementary Fig. 6d) and the wing model can have a spanwise bending, represented by the maximum bending displacement (Supplementary Fig. 6c). An interactive graphic user interface developed using MATLAB (v.7.1, The Mathworks, Inc., Natick, MA, USA) was used to determine the positions and orientations of the body and the wings: the positions and orientations of the models of the body and wings were adjusted until the best overlap between a model image and the displayed frame was achieved in three views, and at this point the positions and orientations of these models were taken as the positions and orientations of the body and the wings. The fitting process was manually done. More detailed description of the method can be found elsewhere[4,23,15].

This process gives the position and orientation of the body, the wing root positions, the Euler angles and the maximum bending displacement of the wings. We processed 360 wingbeats in total, over 20 flight sequences from 20 individuals: 30 wingbeats for each of the 5 biting-midges *Forcipomia gloriose* and for each of the 5 biting-midges *Dasyhelea flaviventris* (these two species have high wingbeat frequency); and 6 wingbeats for each of the 5 gall-midge *Anbremia* sp. and for each of the 5 thrip *Frankliniella occidentalis*.

**Aerodynamic-force computation.** The flows around and the aerodynamic force acting on the insects were computed using the method of computational fluid dynamics (CFD). For medium and large insects at hovering flight, it had been shown that aerodynamic interaction between the body and the wings was negligibly small:

the aerodynamic force in the case with body/wing interaction was less than 2.5% different from that without body/wing interaction[24]. Our computations showed that this also true for the small insects. Therefore, in the present CFD model, only the two wings were considered. The planform of a model wing is approximately the same as that of the corresponding insect wing (the wing planforms for FG, DF, AS and FO are shown in Supplementary Fig. 6d; those for ET, DV, LS and EF in our previous works[4,14,16,15]); the section of the model wing is a flat plat of 3% thickness with rounded leading and trailing edges.

The incompressible Navier–Stokes equations were solved over moving overset grids because there are relative movements between the left and right wings. There was a body-fitted curvilinear grid for each of the wings and a background Cartesian grid which extends to the far-field boundary of the domain (Supplementary Fig. 7a). The flow solver, which was based on an artificial compressibility method developed by Rogers et al.[25], was the same as that used in several previous studies of our group[8,15,19]; its detailed description can be found there.

The solver has been validated by comprehensive tests[15]: comparison with measured data of a revolving wing ($Re=500$)[26] and of a flapping wing ($Re=180$)[27]; comparison with theoretical (Stokes or Oseen solutions) and measured results for a sphere at very low $Re$ (5-20)[28].

Grid resolution tests were conducted to ensure that the flow calculations were grid independent. For the dronefly (ET) wing ($Re=720$), three grid-systems were considered. For grid-system 1, the wing grid had dimensions $41\times 61\times 43$ in the normal direction, around the wing, and in the span-wise direction, respectively (first layer grid thickness was $0.0015c$); the body grid had dimensions $66\times 51\times 35$ along the body, in the azimuthal direction and in the normal direction, respectively; the background grid had dimensions $81\times 81\times 81$ in the $X$, Y and $Z$ directions, respectively. For grid-system 2, the corresponding grid dimensions were $61\times 91\times 65$, $99\times 77\times 53$ and $121\times 121\times 121$ ($0.001c$). For grid-system 3, the corresponding grid dimensions were $91\times 135\times 96$, $149\times 118\times 80$ and $181\times 181\times 181$ ($0.00067c$). For all the three grid-systems, grid points of the background grid concentrated in the near field of the wings where its grid density was approximately the same as that of the outer part of the body-grid. Three grid-systems similar to the above were also used for the small wasp (EF) wing ($Re$=10). The aerodynamic forces computed are shown in Supplementary Fig. 7b and c. It is seen that there is almost no difference between the force coefficients calculated by the three grid-systems. Calculations were also conducted using a larger computational domain. The domain was enlarged by adding more grid points to the outside of the background grid of grid-system 2. The calculated results showed that there was no need to put the outer boundary further than that of grid-system 2. The above results showed that grid-system 2 was proper for the wings of ET and EF. The wings of fruitfly DV, biting midge DF and gall midge AS had similar aspect ratio as that of ET or EF and operated at $Re$ between those of ET and EF, therefore grid-system 2 was also used for insects of these 4 species. For biting midge FG, whose wing had a much larger aspect ratio, the wing grid in grid-system 2 was changed to 65×91×70 (more points in the span-wise direction); and for thrip FO, whose wing had a smaller aspect ratio, the wing grid was changed to 65×91×56 (less points in the span-wise direction). The effect of time step value was also studied and it was found that a numerical solution effectively independent of the time step was achieved if the time step value was ≤ $T/440$, and this value was used in all the calculations.

**Supplementary Information** Supplementary Videos 1-4: near-hover flight of FG1, DF1, AS1 and FO1, respectively. In each video, the left, middle and right parts of the movie show the flight captured by the top-view camera and two side-view cameras, respectively. For FG1, Playback speed is 10fps, approximately 0.1% of the actual speed of the movie; for DF1, 8fps, approximately 0.1%; for AS1 and FO1, 24fps, approximately 0.3%.

**Acknowledgements** This research was supported by grants from the National Natural Science Foundation of China (11832004, 11721202).

**Author Contributions** M.S. conceived the experimental design; Y.Z.L. and H. J. Z. designed and constructed the apparatus and all authors contributed to data collection; Y.Z.L. processed the raw data to extract detailed kinematics and performed the CFD computations; M.S drafted the manuscript; all authors contributed to data interpretation and manuscript preparation.

**Author Information** The authors declare no competing financial interests. Correspondence and requests for materials should be addressed to M.S. (m.sun@buaa.edu.cn)

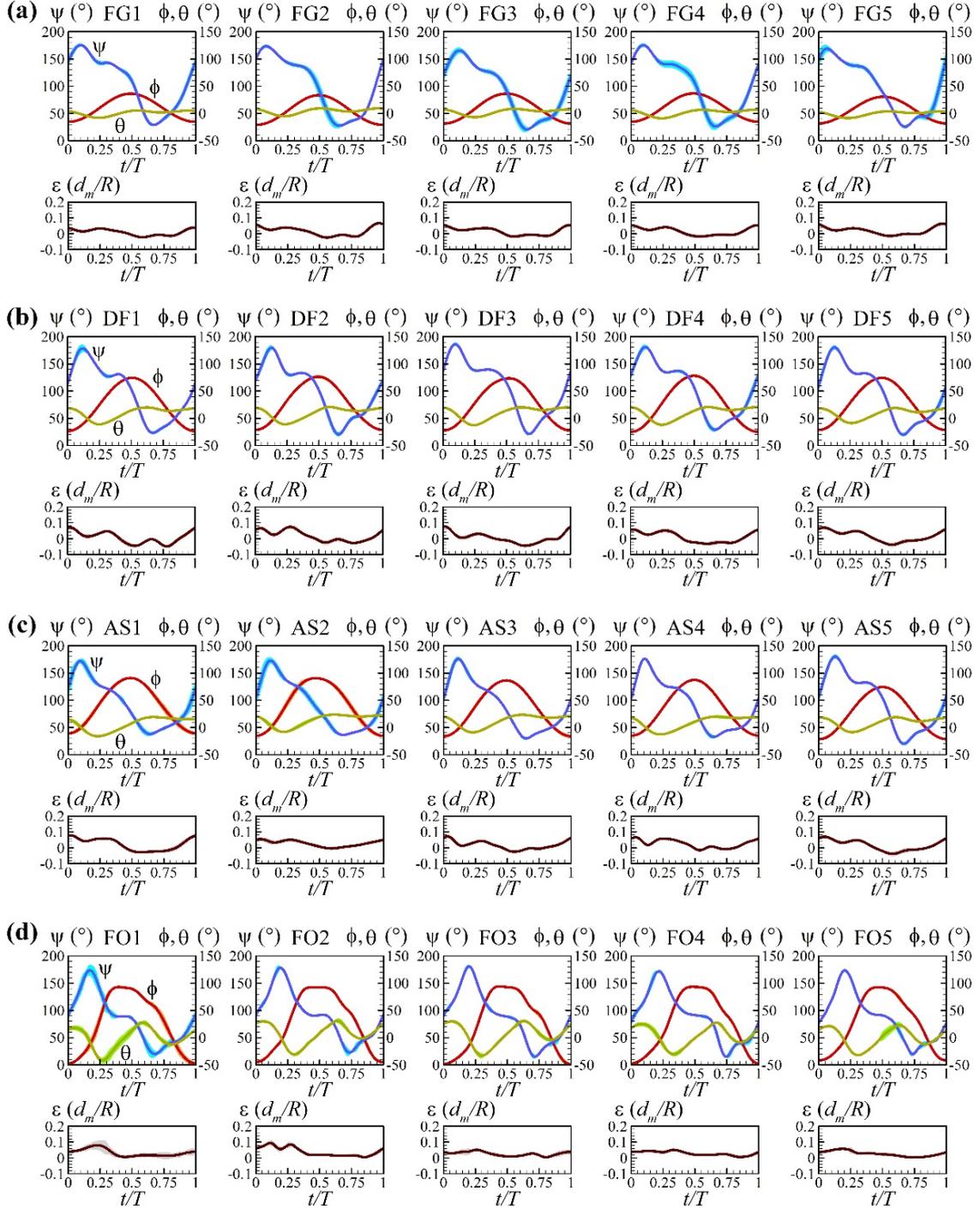

Supplementary Figure 1. Wing kinematics for each of the individuals. Measured Euler angles of wing and the maximum span-wise bending displacement ε ($d_m/R$) in one stroke cycle. **(a)** Biting midge (FG). **(b)** Biting midge (DF). **(c)** Gall midge *Anbremia* sp. (AS). **(d)** thrip *Frankliniella occidentalis* (FO).

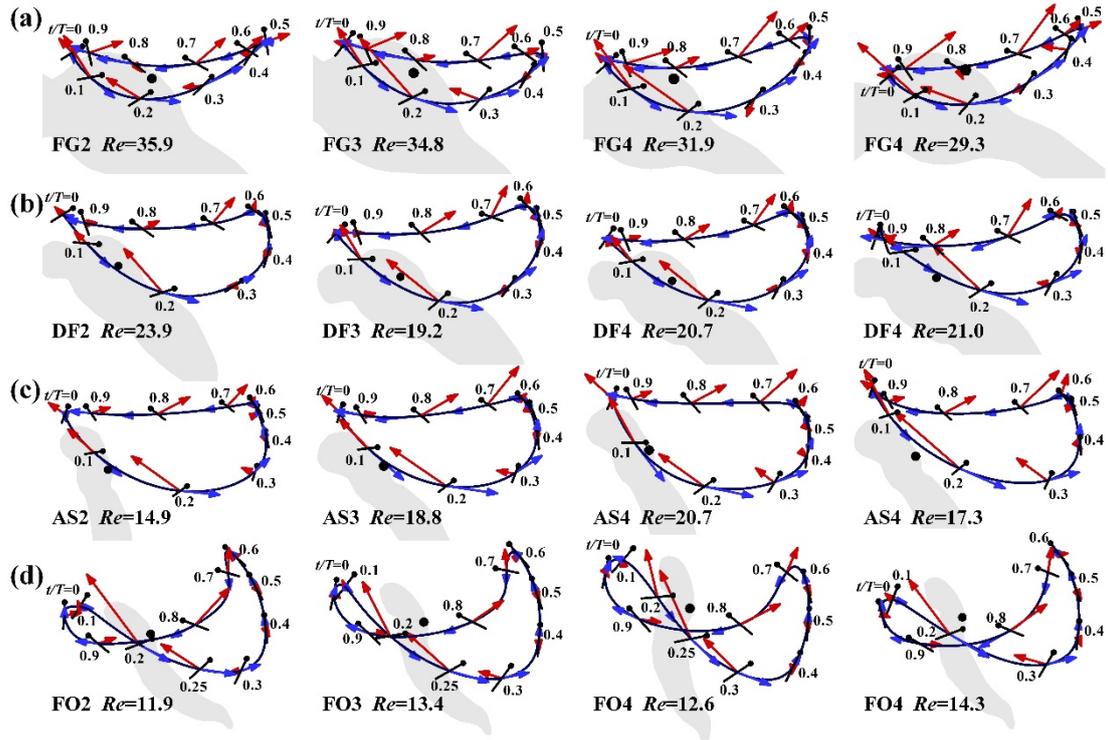

Supplementary Figure 2. Stroke diagrams show the wing motions of the insects. **(a) through (d)**: biting-midge FG, biting-midge DF, gall midge *Anbremia* sp. and thrips *Frankliniella occidentalis*. Solid curve indicates the wing-tip trajectory (projected onto the *x-y* plane); black lines indicate the orientation of the wing at various times in one stroke cycle, with dots marking the leading edge; black dot defines the wing-root location on the insect body; blue arrow, velocity of the wing at the radius of gyration; red arrow, total aerodynamic force of the wing. (Similar plots can be obtained for ET[4], DV[14], LS[16] and EF[15]).

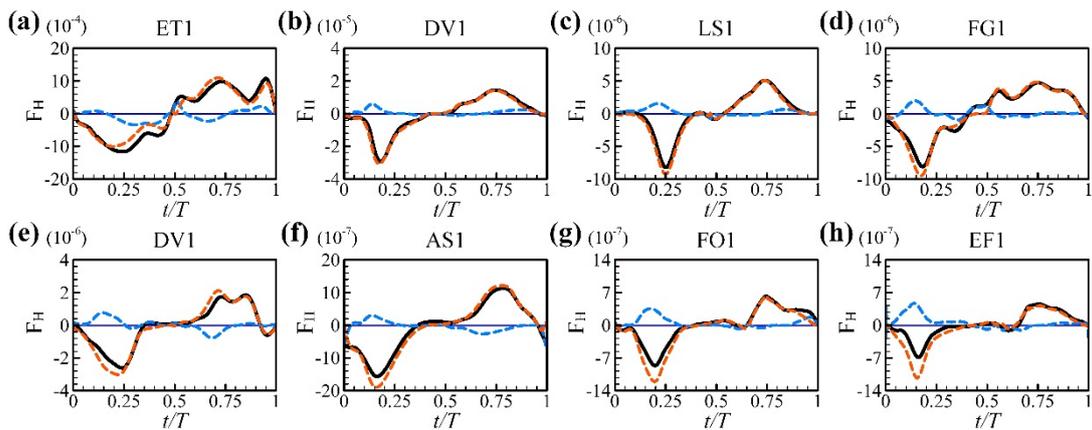

Supplementary Figure 3. Horizontal forces (black line) for the eight insects shown in Fig. 2a-h; contributions by the lift (blue line) and that by drag (orange line). The corresponding vertical forces are in Fig. 2 a*-h*.

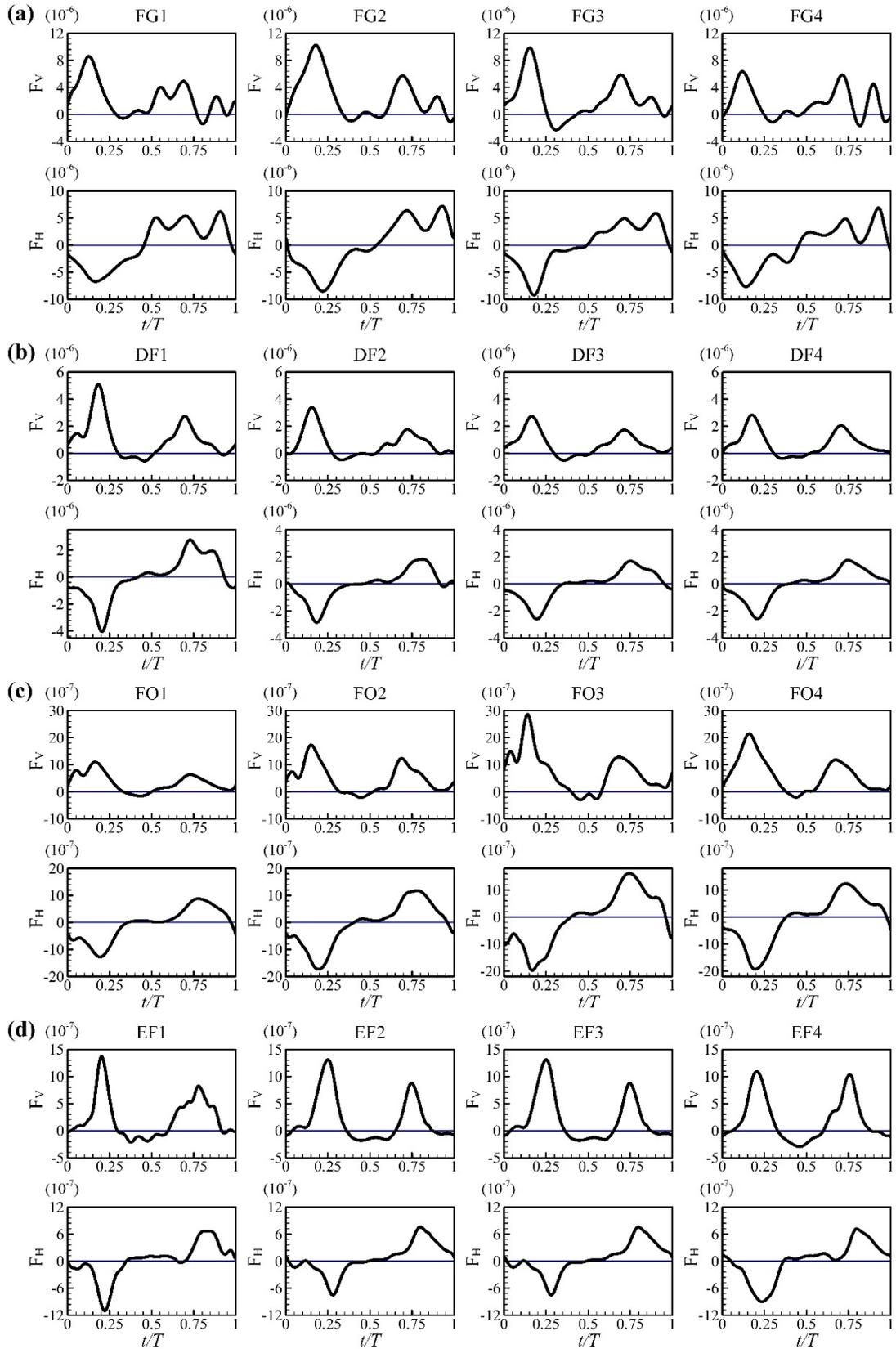

Supplementary Figure 4. Vertical and horizontal forces for the other four individuals of each species.

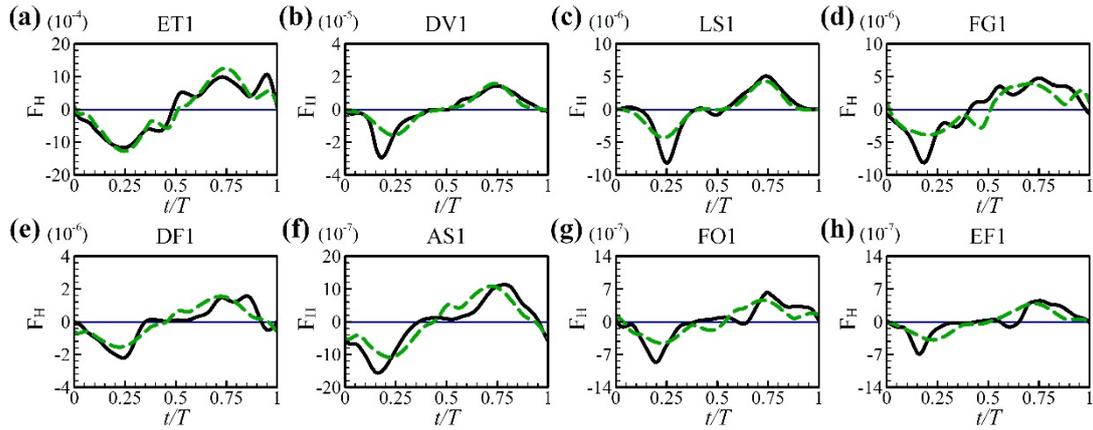

Supplementary Figure 5. Horizontal force produced if the upstroke was planar (green dashed line), compared with that using the real wing kinematics which has U-shape upstroke (black line).

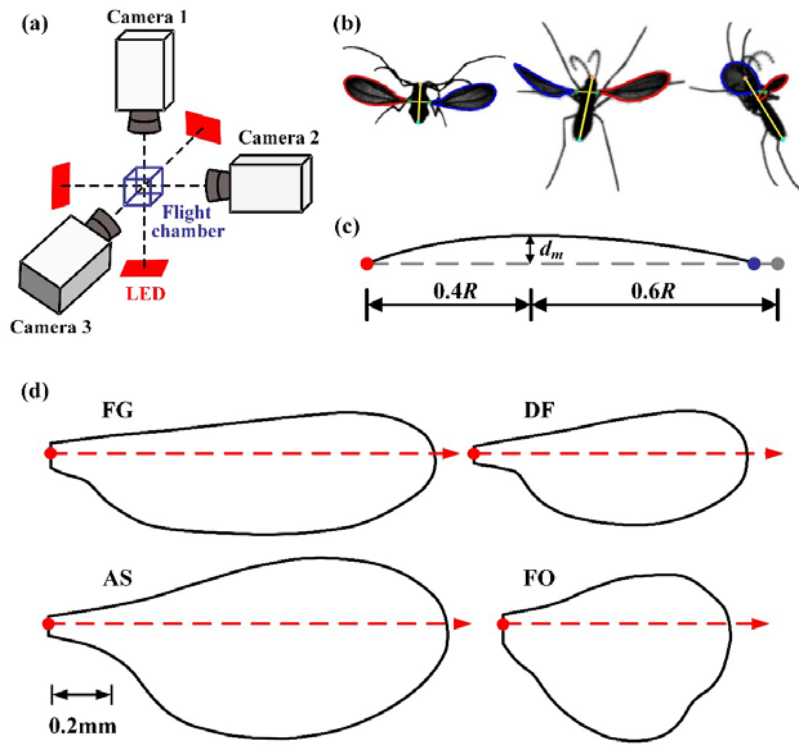

Supplementary Figure 6. Experimental setup, kinematics extraction and wing bending definition. (a) A sketch showing the flight chamber and cameras. Each camera view is backlit using a integrated red light emitting diode (LED). (b) Extraction of body and wing kinematics. (c) Definition of the spanwise bending of wing. $d_m$, maximum bending displacement, assumed to be at $0.4R$. (d) Outline of the wings.

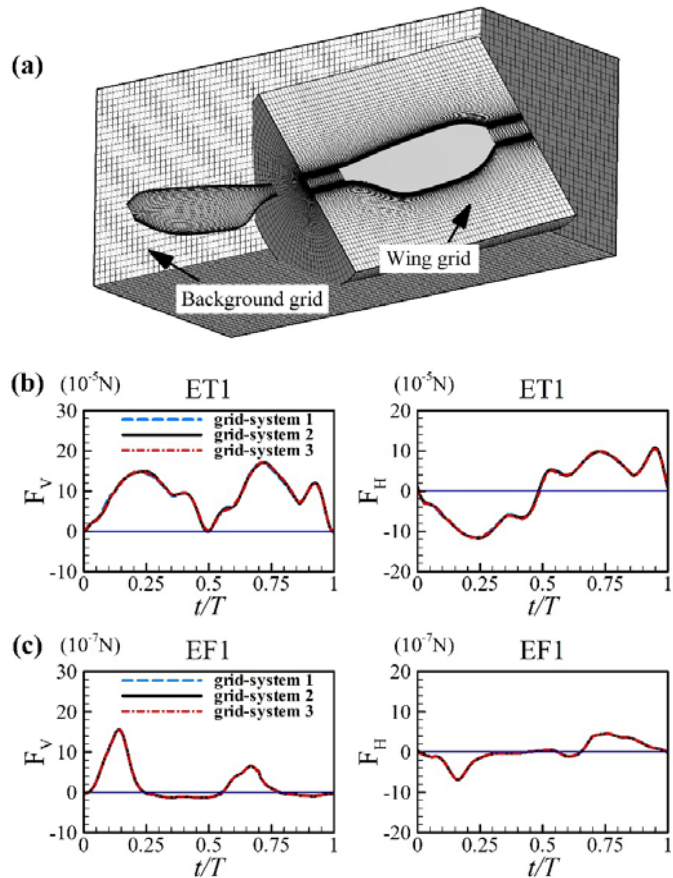

Supplementary Figure 7. (a) Portions of a computational grid-system. (b and c) Time courses of the vertical and horizontal forces in one cycle, calculated with three grid-systems, for ET and EF respectively.

Supplementary Table 1. Flight parameters.

| ID | $Re$ | $R$ (mm) | $S$ (mm$^2$) | $r_2/R$ | $l_r$ (mm) | $l_b$ (mm) | $f$ (Hz) | $\Phi$ (°) | $\beta$ (°) | $u$ (m/s) | $w$ (m/s) | $F_{V,m}$ (μN) | $\eta$ (°) |
|---|---|---|---|---|---|---|---|---|---|---|---|---|---|
| FG1 | 29.8 | 1.30 | 0.38 | 0.60 | 0.53 | 2.29 | 1117 | 52 | 5.3 | 0.00 | 0.00 | 1.849 | 4.5 |
| FG2 | 35.9 | 1.40 | 0.44 | 0.60 | 0.57 | 2.36 | 1112 | 55 | 2.6 | 0.02 | 0.14 | 2.350 | 1.3 |
| FG3 | 34.8 | 1.40 | 0.46 | 0.60 | 0.59 | 2.47 | 1019 | 55 | -4.1 | 0.10 | 0.21 | 2.622 | -1.4 |
| FG4 | 31.9 | 1.34 | 0.41 | 0.60 | 0.55 | 2.33 | 1107 | 52 | 7.2 | 0.12 | 0.07 | 2.216 | 0.9 |
| FG5 | 29.3 | 1.38 | 0.45 | 0.59 | 0.58 | 2.40 | 1016 | 49 | 12.6 | 0.00 | 0.00 | 1.679 | -8.6 |
| DF1 | 23.7 | 0.92 | 0.25 | 0.60 | 0.27 | 1.34 | 723 | 98 | 0.9 | 0.01 | 0.01 | 0.840 | -6.6 |
| DF2 | 23.9 | 0.95 | 0.22 | 0.63 | 0.28 | 1.21 | 781 | 97 | -0.2 | 0.00 | 0.14 | 1.056 | -3.3 |
| DF3 | 19.2 | 0.88 | 0.20 | 0.62 | 0.25 | 1.16 | 724 | 96 | 7.7 | 0.00 | 0.00 | 0.690 | -6.5 |
| DF4 | 20.7 | 0.84 | 0.20 | 0.61 | 0.28 | 1.14 | 739 | 103 | 7.8 | 0.05 | 0.02 | 0.690 | -7.0 |
| DF5 | 21.0 | 0.92 | 0.23 | 0.62 | 0.29 | 1.28 | 698 | 96 | 5.8 | 0.16 | 0.05 | 0.717 | -4.4 |
| AS1 | 17.4 | 1.33 | 0.51 | 0.63 | 0.25 | 1.30 | 238 | 102 | 4.8 | 0.14 | 0.08 | 0.463 | -7.7 |
| AS2 | 14.9 | 1.28 | 0.43 | 0.64 | 0.20 | 1.05 | 230 | 106 | 1.3 | 0.09 | 0.12 | 0.340 | -9.5 |
| AS3 | 18.8 | 1.36 | 0.57 | 0.63 | 0.25 | 1.12 | 230 | 103 | 1.9 | 0.10 | 0.09 | 0.481 | -9.3 |
| AS4 | 20.7 | 1.47 | 0.65 | 0.62 | 0.31 | 1.38 | 226 | 102 | -3.4 | 0.11 | 0.04 | 0.724 | -3.5 |
| AS5 | 17.3 | 1.39 | 0.54 | 0.63 | 0.27 | 1.22 | 213 | 108 | -1.2 | 0.16 | 0.03 | 0.634 | -7.6 |
| FO1 | 13.5 | 0.76 | 0.30 | 0.59 | 0.28 | 1.13 | 239 | 140 | 6.3 | 0.13 | 0.17 | 0.167 | 4.0 |
| FO2 | 11.9 | 0.79 | 0.29 | 0.59 | 0.27 | 1.23 | 223 | 137 | 18.3 | 0.22 | 0.00 | 0.232 | 2.6 |
| FO3 | 13.4 | 0.83 | 0.33 | 0.59 | 0.26 | 1.09 | 213 | 143 | 9.9 | 0.12 | 0.08 | 0.223 | 0.9 |
| FO4 | 12.6 | 0.79 | 0.31 | 0.57 | 0.26 | 1.18 | 225 | 141 | -4.4 | 0.14 | 0.07 | 0.205 | 14.9 |
| FO5 | 14.3 | 0.86 | 0.34 | 0.61 | 0.29 | 1.27 | 231 | 134 | 14.7 | 0.22 | 0.11 | 0.206 | 0.4 |

$Re$, Reynolds number; $R$, wing length; $S$, area of wing; $r_2$, radius of gyration of wing; $l_r$, the distance between the left and right wing-roots; $l_b$, body length. $f$ and $\Phi$, stroke frequency and amplitude, respectively; $\beta$, stroke plane angle; $u$ and $w$, horizontal and vertical velocities of body, respectively; $F_{V,m}$, mean vertical force; $\eta$, angle from the vertical of the mean force vector.